\documentclass[12pt]{article}

\title{Two solvable systems of coagulation equations \\
with limited aggregations} 
\author{Jean Bertoin\thanks{Laboratoire de Probabilit\'es, 
UPMC, 175 rue du Chevaleret, 75013 Paris; and DMA, ENS, 45 rue d'Ulm, 75005 Paris, France. Email:
jean.bertoin@upmc.fr}}

\usepackage{amsfonts,amsmath,amssymb,bbm}
\usepackage{setspace} 
\def\proof{\noindent{\bf Proof:}\hskip10pt}        
\def\QED{\hfill $\Box$}

\font\tenmath=msbm10 scaled 1200
\font\sevenmath=msbm7 scaled 1200
\font\Phiivemath=msbm5 scaled 1200
\newfam\mathfam \textfont\mathfam=\tenmath
\scriptfont\mathfam=\sevenmath \scriptscriptfont\mathfam=\Phiivemath

\def \\ { \cr }
\def\R{\mathbb{R}}
\def \1{1 \mkern -6mu 1} 
\def\N{\mathbb{N}}

\def \p{\frac{\partial}{\partial x}}
\def \e{{\rm e}}
\def \d{\frac{\rm d}{{\rm d} t}}

\newtheorem{theorem}{Theorem}

\newtheorem{lemma}{Lemma}
\newtheorem{corollary}{Corollary}
\begin{document}
\date{}
\maketitle

\begin{abstract}
We consider two simple models for the formation of polymers where at the initial time, each monomer has a certain number of potential links (called arms in the text)
that are consumed when aggregations occur.  Loosely speaking, this imposes restrictions on the number of aggregations. The dynamics of concentrations are governed by modifications of Smoluchowski's coagulation equations.
Applying classical techniques based on generating functions, resolution of quasi-linear PDE's, and Lagrange inversion formula, we obtain explicit solutions to these non-linear systems of ODE's. We also discuss the asymptotic behavior of the solutions and point at some connexions with certain known solutions to Smoluchowski's coagulation equations with
additive or multiplicative kernels.
\vskip 2mm

 \noindent {\bf Keywords : } Coagulation equations, generating function, quasi-linear PDE, Lagrange inversion formula, gelation.
 
 \noindent{\bf A.M.S. Classification } Primary : 34A34 ; Secondary : 82C23, 82D60.

 \end{abstract}

\begin{section}{Introduction}

The coagulation equations of Smoluchowski \cite{Smoluch}  describe the evolution of concentrations of particles in a medium in which particles coalesce pairwise as time passes. The surveys by Aldous \cite{Aldous} and Lauren\c{c}ot and Mischler \cite{LM} provide stimulating introductions to the subject.
We also refer to Drake \cite{Drake} and Dubovski \cite{Dubov} for detailed accounts, and e.g. to \cite{EM, FL, MenPeg} and references therein  for some more recent works in this area. 

In the discrete version of  this model, particles are characterized by an integer $m\geq 1$ which should be though of as a size or a mass, in the sense that the  result  of  the coagulation of a pair of particles, say $\{m,m'\}$,  is $m+m'$. The dynamics are determined in terms  of some symmetric kernel $\kappa: \N^*\times \N^*\to \R_+$ such that, loosely speaking, every  pair of particles $\{m,m'\}$ coalesces at
rate  $\kappa(m,m')$. This means that if
 $c_t(m)$ denotes  the concentration of particles with size $m$ at time $t$, then
the evolution of concentrations is  governed by the infinite system
\begin{equation}\label{EQ1}
\d c_t(m)=\frac{1}{2}\sum_{m'=1}^{m-1} 
c_t(m')c_t(m-m')\kappa(m',m-m')
-c_t(m)\sum_{m'=1}^{\infty}c_t(m')\kappa(m,m')\,,
\end{equation}
where the first term in the right-hand side accounts for the creation of particles of size $m$ by the coagulation of a pair of particles $\{m',m-m'\}$, the factor $\frac{1}{2}$ stems from an obvious symmetry, and the second term accounts for the disappearance of particles with size $m$ as the result of a coagulation with other particles.

In general \eqref{EQ1} cannot be solved explicitly; however there are a few important exceptions.  The fundamental  examples of solvable Smoluchowski's equations occur when the coagulation  kernel $\kappa$ is constant, additive, or multiplicative.  For instance, for a monodisperse initial condition (this means that at time $t=0$, all particles have unit size), 
it is known since the original work of Smoluchowski \cite{Smoluch} that
\begin{equation}\label{EQ2}
c_t(m) = \left(1+\frac{t}{2}\right)^{-2}\left(\frac{t}{2+t}\right)^{m-1}\,,\qquad 0\leq t < \infty
\end{equation}
solves \eqref{EQ1} for $\kappa(m,m')=1$.
Further, if 
$$B(\lambda,m)= (\lambda m)^{m-1}\e^{-\lambda m}/m!\,,\qquad m\in\N^*$$ 
denotes the Borel probability function with parameter $\lambda\in[0,1]$, then, according to Golovin \cite{Golov},
\begin{equation}\label{EQ3}
c_t(m)=\e^{-t}B(1-\e^{-t},m)\,,\qquad 0\leq t < \infty
\end{equation}
is a solution to \eqref{EQ1} for $\kappa(m,m')=m+m'$. Finally,  it has been shown by McLeod \cite{McLeod} that
\begin{equation}\label{EQ4}
c_t(m)=m^{-1}B(t,m)\,,\qquad 0\leq t < 1 
\end{equation}
solves \eqref{EQ1} for $\kappa(m,m')=mm'$.
More generally, explicit though complicated expressions can be obtained for solutions to \eqref{EQ1} when the coagulation kernel $\kappa$ is a linear combination of these three basic kernels; see \cite{Trub, Spouge, vDE1}.

The key reason why Smoluchowski's equations can be solved explicitly in these cases  is that considering generating functions transforms \eqref{EQ1} into a solvable  PDE; see e.g. Deaconu and Tanr\'e \cite{DT}.
The purpose of the present work is to point out that the same techniques can be applied successfully to investigate a related but more complex model for coagulation.
Roughly speaking, the dynamics described by
Smoluchowski's equations depend only on an additive quantity, namely the size of particles. Norris  \cite{Norris}  developed the much more sophisticated setting of cluster coagulation models in which the rates of transition may depend on
further parameters, such as typically
the shapes of the clusters of particles. In this direction,  we will deal here with  a toy model for the formation of polymers in which
coagulation rates are functions of a non-additive quantity that depends on the history of polymers. More precisely,  the medium consists initially in
monomers (elementary particles with unit size), such that each monomer has a certain number of potential links which we call  {\it arms}. Specifically, we suppose that each particle is determined by a pair of integers $(a,m)$, where $a\geq 0$ is the number of  arms of the particle and $m\geq 1$ represents its size. The arms serve to perform aggregations, i.e. to connect pairs of particles, and are consumed each time an aggregation occurs. We will consider two different models.

 The first model will be studied in Section 2. It is {\it oriented}, in the sense that a coagulation occurs when a particle uses one of its arms to grab another particle. Only one arm is consumed for each coagulation event; specifically, 
when a particle $(a,m)$ with $a\geq 1$ grabs a particle $(a',m')$, these two particles merge into a single one $(a+a'-1, m+m')$. We assume further that each arm grabs other particles at uniform rate; the precise formulation will be given in Section 2.1. In Section 2.2 we use generating functions to connect these dynamics  to a quasi-linear PDE that is then solved by the method of characteristics. Section 2.3 presents a version of Lagrange inversion formula which will be needed to invert the generating functions. The main results on the oriented model are given in Section 2.4, and finally some illustrative examples are discussed in Section 2.5. In particular, we point at  some resemblances between on the one hand,
the oriented model when the number of arms of monomers is distributed according to the standard Poisson law, and on the other hand, Smoluchowski's coagulation equation for the additive kernel.

The second model will be studied in Section 3. It is {\it symmetric}, in the sense that  each coagulation consumes one arm for each of the two particles involved into a coagulation event~: 
when a pair of particles $\{(a,m),(a',m')\}$ with $aa'\geq 1$ coalesces, the resulting particle is \linebreak  $(a+a'-2, m+m')$.  Furthermore, we assume that each pair of arms is activated uniformly. Section 3.1 introduces the setting rigorously, and points at a critical time $\Gamma_{\infty}$ in the evolution of the system which resembles the gelation time (i.e. the instant when particles with infinite size appear in the medium)
 in Smoluchowski's coagulation equations. We also use generating functions to make the connexion with a quasi-linear PDE closely related to the one that arises for the oriented model. The main results and some  examples are given in Section 3.2.  For instance, we determine the critical time $\Gamma_{\infty}$ in terms of the initial data (interestingly,  $\Gamma_{\infty}$ may be finite or infinite) and note some similarities between the symmetric model started from monomers with arms distributed according to the standard Poisson law and McLeod's solution to Smoluchowski's coagulation equation for the multiplicative kernel at gelation time.

We  conclude this Introduction by explaning the term {\it limited} in the title. Roughly speaking, the effect 
of introducing arms in a coagulation model is that it imposes some restrictions to the number of aggregations. Indeed, in the oriented model, we shall show that if at the initial time the mean number of arms is less than the total concentration of particles, then the system converges as time tends to infinity to some limiting concentrations (and more precisely the system does not create particles with infinite size); see Corollary \ref{C1}. A similar phenomenon is observed for the symmetric model; see Corollary \ref{C2} and the remark thereafter. Perhaps this is more surprising for the symmetric model, as the latter bears some similarities with the classical Smoluchowski's coagulation for the multiplicative kernel,
and it is well-known that gelation  then always occurs in that case.

\end{section}

\begin{section}{The oriented model}
\subsection{Setting}
We now describe precisely the dynamics for oriented coagulation with arms. For every $t\geq 0$, 
$a\in\N:=\{0,1,2,\ldots\}$ and $m\in\N^*:=\{1,2,\ldots\}$, let $c_t(a,m)$ denote the density of particles $(a,m)$ at time $t$. Loosely speaking, we suppose that at any time, each arm may grab each particle at the same unit rate. Rigorously, this means in terms of concentrations that the transition
\begin{equation}\label{EQ4'}
\{(a,m),(a',m')\}\longrightarrow (a+a'-1,m+m')
\end{equation}
occurs at time $t$ with intensity
$$ac_t(a,m)c_t(a',m') + a'c_t(a',m')c_t(a,m)=(a+a')c_t(a,m)c_t(a',m').$$
The evolution of the concentration functions is thus governed by an infinite non-linear system of ODE's
\begin{eqnarray}\label{EQ5}
\d c_t(a,m)
 \ =&&\frac{1}{2}\sum_{a'=0}^{a+1}\sum_{m'=1}^{m-1} (a+1)
c_t(a',m')c_t(a-a'+1,m-m')\nonumber \\
&&-c_t(a,m)\sum_{a'=0}^{\infty}\sum_{m'=1}^{\infty}(a+a')c_t(a',m')\,.
\end{eqnarray}

We point out that
$$
 \frac{1}{2}\sum_{a'=0}^{a+1} \sum_{m'=1}^{m-1}(a+1)
c_t(a',m')c_t(a-a'+1,m-m') 
= \sum_{a'=0}^{a+1} \sum_{m'=1}^{m-1} a'
c_t(a',m')c_t(a-a'+1,m-m')\,,$$
and since in the sum in the right hand side, the term corresponding to $a'=0$ is null,  
we can reformulate \eqref{EQ5} as
$$\d c_t(a,m)=\sum_{a'=1}^{a+1}\sum_{m'=1}^{m-1} a'
c_t(a',m')c_t(a-a'+1,m-m')
-c_t(a,m)\sum_{a'=0}^{\infty}\sum_{m'=1}^{\infty}(a+a')c_t(a',m')\,.
$$
 The latter expression may be closer to the asymmetric description of the dynamics 
of the system.

We also introduce the notation
$$C_t:=\sum_{a=0}^{\infty}\sum_{m=1}^{\infty}c_t(a,m)$$
for the total concentration of particles at time $t$, and
$$A_t:=\sum_{a=0}^{\infty}\sum_{m=1}^{\infty}ac_t(a,m)$$
for the mean number of arms, and stress that \eqref{EQ5} only makes sense as long as
$A_t$ and $C_t$ are finite. But it is easy to check  (see Lemma \ref{L0} below) that it suffices to assume that at the initial time $A_0$ and $C_0$ are finite.

It will be convenient to re-express the system \eqref{EQ5} in a slightly different form by introducing the notation 
$$\langle c_t,f\rangle := \sum_{a=0}^{\infty}\sum_{m=1}^{\infty} f(a,m)c_t(a,m)\,,$$
where $f: \N\times \N^* \to \R_+$ stands for a generic  nonnegative function. 

\begin{lemma}\label{L0} For any solution to \eqref{EQ5}, the function
$t\to A_t+C_t$ is non-increasing; as a consequence $A_t+C_t\leq A_0+C_0$ for all $t\geq 0$. Further we have 
\begin{equation}\label{EQ6}
\d \langle c_t,f\rangle=\frac{1}{2}
\sum_{a,a'=0}^{\infty}\sum_{m,m'=1}^{\infty} (f(a+a'-1,m+m')- f(a,m)-f(a',m'))(a+a')
c_t(a,m)c_t(a',m')
\end{equation}
whenever $f: \N\times \N^* \to \R$ is bounded.
\end{lemma}

\proof First observe that
 \eqref{EQ5} is precisely \eqref{EQ6}  when $f$ is a Dirac function. By linearity, \eqref{EQ6} thus holds when $f$ has finite support.
 
Next fix $n\in\N^*$ and define
 $$\varphi_n(a):=\left\{ \begin{matrix}
1+a &\hbox{ if }1+a\leq n, \\
2n-1-a &\hbox{ if }n\leq 1+a\leq 2n, \\
0 &\hbox{ otherwise.}\\
\end{matrix}\right.
 $$
Then consider the function with finite support $f_n(a,m):={\bf 1}_{\{m\leq n\}}\varphi_n(a)$. It is readily checked that 
$$f_n(a+a'-1,m+m')- f_n(a,m)-f_n(a',m')\leq 0$$
for all $ a,a'\in\N $ and $
m,m'\in\N^*$. We deduce from \eqref{EQ6} that $t\to  \langle c_t,f_n\rangle$ is non-increasing. Further, $f_n(a,m)$ increases to $1+a$ as $n\uparrow\infty$, so by
 monotone convergence we have that $A_t+C_t=\lim_{n\uparrow \infty} \langle c_t,f_n\rangle$, and we conclude that $t\to A_t+C_t$ is non-increasing.

As we already know that \eqref{EQ6} holds when $f$ has finite support, a standard approximation procedure yields the extension when 
$f$ is only bounded by dominated convergence. \QED

We now start the analysis by pointing at an elementary property which should be intuitively obvious, as at each coagulation event, the number of particles and the total number of arms both decrease by one unit. The rigorous proof requires however some care.

\begin{lemma}\label{L1} For any solution to \eqref{EQ5}, the difference $D:=C_t-A_t$ remains constant as time passes.
More precisely :

\noindent{\rm (i)}  If $C_0=A_0$, then 
$$C_t=A_t=C_0/(1+tC_0)\,.$$

\noindent{\rm (ii)}  If $C_0-A_0={D} \neq0$, then 
$$C_t=A_t+{D} = {D} \frac{C_0 \e^{{D} t}}{C_0
(\e^{{D} t}-1)+{D}}\,.$$
\end{lemma} 

\proof We get from \eqref{EQ6} for $f\equiv 1$ that
$\d C_t=-A_tC_t$, and in particular the total concentration decreases with $t$.
Similarly, fix $\alpha>0$ and apply \eqref{EQ6} for $f^{(\alpha)}(a,m)=\alpha\wedge a$
to see that $A^{(\alpha)}_t:=\langle c_t,f^{(\alpha)}\rangle$ fulfills $\d A^{(\alpha)}_t\leq -A_tC_t $. As $A_t=\lim_{\alpha\uparrow \infty}\uparrow A^{(\alpha)}_t$, this entails that
$A_t$ also decreases with $t$. 

Then observe that
$$|f^{(\alpha)}(a+a'-1,m+m')-f^{(\alpha)}(a,m)-f^{(\alpha)}(a',m')|\leq (a\wedge a')+1\,,$$
and that therefore
\begin{equation}\label{EQ7}
|f^{(\alpha)}(a+a'-1,m+m')-f^{(\alpha)}(a,m)-f^{(\alpha)}(a',m')|(a+a')\leq 2 aa'+a+a'\,.
\end{equation}
Since
$$\lim_{\alpha\to\infty}(f^{(\alpha)}(a+a'-1,m+m')-f^{(\alpha)}(a,m)-f^{(\alpha)}(a',m'))
=-1\,,$$
it is easy to get from \eqref{EQ6} by dominated convergence that $\d A_t=-A_tC_t$.
We conclude that the difference $C_t-A_t$ is constant and 
the explicit expressions in the statement follow readily. \QED

\subsection{A quasi-linear PDE}
 The system of evolution equations \eqref{EQ6} resembles Smoluchowski's coagulation equation for the additive kernel. In the latter case, it is well-known that considering exponential functions yields a quasi-linear PDE related to the transport equation which can be solved explicitly, see e.g. \cite{DT}. 
This invites us to introduce the generating functions 
$$g_t(x,y):=\langle c_t,f_{x,y}\rangle \hbox{ with } f_{x,y}: (a,m)\to x^a y^m
\hbox{ and }x,y\in[0,1]\,.$$

For the sake of simplicity, we will focus on the situation where the total concentration of particles equals $1$  at the initial time, i.e. $C_0=1$. The general case can easily be reduced to that one by a linear time-substitution. 

\begin{lemma}\label{L2} {\rm (i)} Consider initial concentrations $(c_0(a,m): a\in\N\hbox{ and }m\in\N^*)$ such that  $C_0=A_0=1$. The system \eqref{EQ5} is then equivalent to 
the quasi-linear PDE
$$\d g_t(x,y)=\left(g_t(x,y)-\frac{x}{1+t}\right)\p g_t(x,y)-\frac{1}{1+t}g_t(x,y)$$
 for the generating functions, where $t\geq 0$ and $x,y\in[0,1]$. The latter possesses a unique solution which is given in terms of  its value at boundary $t=0$ by
 $$g_t(x,y) = (1+t)^{-1}g_0(h_t(x,y),y)= t^{-1} h_t(x,y)-\frac{x}{t^2+t}\,,$$
where $h_t(\cdot,y): [0,1]\to [0,1]$ is the (unique) inverse of the function $x\to (1+t)x-tg_0(x,y)$.

\noindent {\rm (ii)} Consider initial concentrations $(c_0(a,m): a\in\N\hbox{ and }m\in\N^*)$ such that $C_0=1$ and $C_0-A_0={D}$ for some ${D}\in
(-\infty,1)\backslash\{0\}$. The system \eqref{EQ5} is then equivalent to 
the quasi-linear PDE
$$\d g_t(x,y)=\left(g_t(x,y)-x\frac{ {D} \e^{{D} t}}{\e^{{D} t}-1+{D}}\right)
\p g_t(x,y)+\left({D}- \frac{ {D} \e^{{D} t}}{\e^{{D} t}-1+{D}}
\right)g_t(x,y) $$
 for the generating functions, where $t\geq 0$ and $x,y\in[0,1]$. The latter possesses a unique solution which is given in terms of  its value at boundary $t=0$ by
  \begin{eqnarray*}
g_t(x,y) &=& \frac{ {D} \e^{{D} t}}{\e^{{D} t}-1+{D}}
g_0(h_t(x,y),y) \\
&=&  \frac{ {D} \e^{{D} t}}{\e^{{D} t}-1} h_t(x,y)- \frac{ {D}^2 \e^{{D} t}}
{(\e^{{D} t}-1+{D})(\e^{{D} t}-1)}\, x\,,
\end{eqnarray*}
where $h_t(\cdot,y): [0,1]\to [0,1]$ is the (unique) inverse of the function
$$x\to {D}^{-1}\left(\left(\e^{{D} t}-1+{D}\right)x-\left(\e^{{D} t}-1\right)g_0(x,y)\right)\,.$$
\end{lemma}

\proof (i) From \eqref{EQ6} for $f=f_{x,y}$, we get 
$$\d g_t(x,y)= (\alpha-\beta-\gamma)/2$$
where $\alpha$ is given by 
\begin{eqnarray*}
& &
\sum_{a,a'=0}^{\infty}\sum_{m,m'=1}^{\infty} ax^{a-1}y^m
c_t(a,m)x^{a'} y^{m'}c_t(a',m')
+ \sum_{a,a'=0}^{\infty}\sum_{m,m'=1}^{\infty} a'x^{a'-1}y^{m'}
c_t(a',m') x^{a} y^{m} c_t(a,m)\\
&=& 2 g_t(x,y) \p g_t(x,y)\,,
\end{eqnarray*}
and $\beta=\gamma$ by 
\begin{eqnarray*}
& &x
\sum_{a,a'=0}^{\infty}\sum_{m,m'=1}^{\infty} ax^{a-1}y^m
c_t(a,m)c_t(a',m')
+ \sum_{a,a'=0}^{\infty}\sum_{m,m'=1}^{\infty} a'x^{a}y^{m}
 c_t(a,m) c_t(a',m')\\
&=& x C_t\p g_t(x,y) + g_t(x,y) A_t\,.
\end{eqnarray*}
Recall from Lemma \ref{L1} that
$C_t=A_t=1/(1+t)$ to get the PDE in the statement.

Note that this PDE does not involve partial derivatives with respect to the variable $y$, and is quasi-linear. It can be solved using the methods of characteristics.
Specifically we consider functions $t=t(r,s)$, $x=x(r,s)$ and $u=u(r,s)$ for $0\leq r \leq   1$ and $s\geq 0$, with the boundary conditions $t(r,0)=0$, $x(r,0)=r$, $u(r,0)=g_0(r,y)$ and
such that
$$\frac{{\rm d}t}{{\rm d}s}=1\quad,\quad \frac{{\rm d}x}{{\rm d}s}= \frac{x}{1+t}-u\quad,\quad
\frac{{\rm d}u}{{\rm d}s}=-\frac{u}{1+t}\,.$$
We deduce
 $$t(r,s)=s\quad,\quad u(r,s)=g_0(r,y)/(1+s)\quad,\quad x(r,s)=r(1+s)-sg_0(r,y)\,;$$
and then the comparison with the quasi-linear PDE gives $g_t(x,y)=u(r,s)$.

The requirements $C_0=g_0(1,1)=1$ and $A_0=\p g_0(1,1)=1$ ensure that for any $s\geq 0$ and $y\in[0,1]$, the
function $r\to r(1+s)-sg_0(r,y)$ has derivative $1+s(1-\p g_0(r,y))\geq 1$, and since 
$-sg_0(0,y)<0<1\leq (1+s)-sg_0(1,y)$, there is a unique solution $h_s(\cdot,y):[0,1]\to[0,1]$ to the equation
$h_s(x,y)=h_s(x,y)(1+s)-sg_0(h_s(x,y),y)$. This provides the solution given in the statement. 

 Conversely, if $(g_t: t\geq0)$ is a family of  generating functions solving this PDE with a boundary condition $g_0(x,y)=\langle c_0,f_{x,y}\rangle$ and such that
$g_0(1,1)=C_0=1$ and $\p g_0(1,1)=A_0=1$, then we know from above that
 $$g_t(x,y) = (1+t)^{-1}g_0(h_t(x,y),y)= t^{-1} h_t(x,y)-\frac{x}{t^2+t}\,,$$
with $h_t(\cdot,y): [0,1]\to [0,1]$ the (unique) inverse of  $x\to (1+t)x-tg_0(x,y)$. 
Note that $h_t(1,1)=1$ for all $t$, which gives $g_t(1,1)=1/(1+t)$. Further,
taking a (left) derivative with respect to the variable $x$ at $x=1$ in the identity
$$x= (1+t)h_t(x,1)-tg_0(h_t(x,1),1)$$
entails $\p h_t(1,1)=1$ and then 
$$\p g_t(1,1)=t^{-1}\p h_t(1,1)-1/(t^2+t)=1/(1+t)\,.$$
Thus the family $(g_t: t\geq 0)$ also solves the equation
$$\d g_t(x,y)=\left(g_t(x,y)-x g_t(1,1)\right)\p g_t(x,y)-g_t(x,y)\p g_t(1,1)\,.$$
Since $g_t(1,1)=\langle c_t,1\rangle=C_t$ and $\p g_t(1,1)=\langle c_t,a\rangle=A_t$ , we recover \eqref{EQ5} by inverting the generating functions.

(ii) The derivation of the PDE is obtained as in the proof of (i), applying
the second part of Lemma \ref{L1} in place of the first. 
It is solved again by the methods of characteristics; using the same notation as in (i), we are led to consider the system of ODE's
$$\frac{{\rm d}t}{{\rm d}s}=1\quad,\quad \frac{{\rm d}x}{{\rm d}s}= x\,\frac{ {D} \e^{{D} t}}{\e^{{D} t}-1+{D}}-u\quad,\quad
\frac{{\rm d}u}{{\rm d}s}=D\left(1-\frac{  \e^{{D} t}}{\e^{{D} t}-1+{D}}\right) u$$
 with the boundary conditions $t(r,0)=0$, $x(r,0)=r$, $u(r,0)=g_0(r,y)$.
We obtain
 $$t(r,s)=s\quad,\quad u(r,s)=g_0(r,y) \frac{ {D} \e^{{D} s}}{\e^{{D} s}-1+{D}}$$
 and 
 $$ x(r,s)=D^{-1}\left(r\left(\e^{{D} s}-1+{D}\right)
 -g_0(r,y)\left( \e^{{D} s}-1\right)\right)\,.$$
The proof can then be completed just as in the case (i). \QED

\subsection{A version of Lagrange inversion formula}
The final step of our analysis consists in checking that the function $(x,y)\to h_t(x,y)$ that appears in Lemma \ref{L2}
is the generating function of some finite measure on $\N\times \N^*$
that can be inverted explicitly, at least under some natural hypotheses. It is an easy application of Lagrange inversion formula (see, e.g. Section 5.1 in \cite{Wilf}).

\begin{lemma}\label{L3} Let $\mu=(\mu(a))_{a\in\N}$ be a finite measure on $\N$, $\mu\not\equiv 0$. We write
$$g(x)=\sum_{a=0}^{\infty} x^a\mu(a)\,,\qquad x\in[0,1]$$
for its generating function. For every $p,q>0$ and for every $x,y>0$ sufficiently small,
the equation 
$$h(x,y)=yg(px +qh(x,y))$$
has a unique solution which  is analytic in $x$ and $y$ and given by
$$h(x,y)=\sum_{a=0}^{\infty}\sum_{m=1}^{\infty}x^{a}y^m\frac{1}{m} \left(  \begin{matrix} m+a-1\\ a\\
\end{matrix}\right)q^{m-1}p^{a}\mu^{*m}(m+a-1)\,,$$
where  $\mu^{*m}$ stands for the $m$-th convolution power of $\mu$.
\end{lemma}

\proof  We fix $p,q>0$ and $x$ sufficiently small so that $px<1$. Then we define $\tilde g(y):=g(px+qy)$
 for $y\geq 0$ with $px+qy\leq 1$, which is the generating function of the sigma-finite measure
$$\tilde \mu(k):=\sum_{n=k}^{\infty} \left(  \begin{matrix} n\\ k\\
\end{matrix}\right) (px)^{n-k} q^k\mu(n)\,,\qquad k\in\N\,. $$
Observe also that $\tilde g(0)=g(px)>0$. 
According to the Lagrange inversion formula, the equation $h(x,y)=y\tilde g(h(x,y))$
has a unique solution for $y>0$ sufficiently small, which is analytic in $y$ and can be expressed as
$$h(x,y)=\sum_{n=1}^{\infty}y^n n^{-1}\tilde \mu^{*n}(n-1)\,.$$
Then observe that the generating function of the  measure $\tilde \mu^{*n}$
is $\tilde g^n(y)=g^n(px+qy)$, so
$$\tilde \mu^{*n}(k)=\sum_{j=k}^{\infty} \left(  \begin{matrix} j\\ k\\
\end{matrix}\right) (px)^{j-k} q^k\mu^{*n}(j)\,. $$
We deduce that
$$
h(x,y)=\sum_{n=1}^{\infty}y^n n^{-1} \sum_{j=n-1}^{\infty} \left(  \begin{matrix} j\\ n-1\\
\end{matrix}\right) (px)^{j-n+1} q^{n-1}\mu^{*n}(j)\,,
$$
and the change of variables $j=a+n-1$ completes the proof. \QED

\subsection{Explicit solutions}
We are now able to solve \eqref{EQ5} explicitly in the situation when the initial concentrations $c_0(a,m)$ are carried by particles having unit size; recall also
the assumption that the initial total concentration is $1$. In this direction, it may be interesting to point out that even though the number of arms of a particle is not
an additive quantity, the simple change of variables $\alpha=a-1$ yields a new parametrization of particles that is additive on coagulation, in the sense that
\eqref{EQ4'} then reads
\begin{equation}\label{EQ4"}
\{(\alpha,m),(\alpha',m)\}\longrightarrow (\alpha+\alpha', m+m')\,.
\end{equation}
This observation makes the solvability of \eqref{EQ5} by techniques based on generating functions easier to understand. Note however that the parameter $\alpha$ may take the
value $-1$ and that the rate of the coagulation \eqref{EQ4"} is now
$a+a'=\alpha+\alpha'+2$ and hence not additive in the alternative parameter $\alpha$.

More precisely,  
we consider $\mu=(\mu(a))_{a\in\N}$ a probability measure on $\N$, and  
denote for every $m\in\N^*$ by $\mu^{*m}$ the $m$-th convolution power of $\mu$. 
 
\begin{theorem}\label{T1} The system \eqref{EQ5} has a unique solution 
$(c_t(a,m): a\in\N, m\in\N^* \hbox{ and } t\geq0)$  started from
$$c_0(a,m)={\bf 1}_{\{m=1\}}\mu(a)\,,\qquad a\in\N \hbox{ and } m\in\N^*$$
which is given as follows :

\noindent {\rm (i)} If $\sum_{a\in\N}a\mu(a)=1$,
then 
$$c_t(a,m)= m^{-1}t^{m-1}(1+t)^{-(a+m)} \left(  \begin{matrix} a+m-1\\ a\\
\end{matrix}\right) \mu^{*m}(a+m-1)\,.$$
\noindent {\rm (ii)} If $\sum_{a\in\N}a\mu(a)=1-D$
for some $D\in(-\infty,1)\backslash\{0\}$, 
then
$$c_t(a,m)= \e^{Dt}m^{-1} D^{a+1}\left(\e^{Dt}-1\right)^{m-1}
\left(\e^{Dt}-1+D\right)^{-(a+m)} \left(  \begin{matrix} a+m-1\\ a\\
\end{matrix}\right) \mu^{*m}(a+m-1)\,.$$
\end{theorem}

\proof We shall focus on (i), the argument for (ii) being similar
(note also that taking the limit as $D\to0$
in part (ii) yields the solution in the case (i), as it should be expected).

According to Lemma \ref{L2} and the present hypotheses, 
we have to consider the initial generating function 
$$g_0(x,y)=\langle c_0, f_{x,y}\rangle =\sum_{a=0}^{\infty} x^a y \mu(a)=yg(x)$$
where $g$ is as in Lemma \ref{L3}, and then the solution $h_t(x,y)$ to
$$(1+t)h_t(x,y)-tyg(h_t(x,y))=x\,.$$
Lemma \ref{L3} invites us to introduce 
$$h(x,y):=t^{-1}((1+t)h_t(x,y)-x)\,,$$
so that 
$$h(x,y)=yg\left(\frac{x}{1+t} +\frac{t}{1+t}h(x,y)\right)\,.$$
According to Lemma \ref{L3} (with $p=1/(1+t)$ and $q=t/(1+t)$), the solution to this equation can be expressed in the form
$$h(x,y)=\sum_{a=0}^{\infty}\sum_{m=1}^{\infty}x^{a}y^m\frac{1}{m} \left(  \begin{matrix} m+a-1\\ a\\
\end{matrix}\right)\left(\frac{t}{1+t}\right)^{m-1}(1+t)^{-a}\mu^{*m}(m+a-1)\,,$$
whenever $x,y$ are sufficiently small. 
Recall that
$$h_t(x,y)=(1+t)^{-1}(th(x,y)+x)\,,$$
so we get
$$h_t(x,y)=\frac{x}{1+t}+\sum_{a=0}^{\infty}\sum_{m=1}^{\infty}x^{a}y^m\frac{1}{m} \left(  \begin{matrix} m+a-1\\ a\\
\end{matrix}\right)t^m(1+t)^{-(a+m)}\mu^{*m}(m+a-1)\,.$$

We then known that the unique solution to the quasi-linear PDE in Lemma \ref{L2}(i)  is\begin{eqnarray*}
g_t(x,y)&= &t^{-1} h_t(x,y)-\frac{x}{t^2+t}\\
&=&\sum_{a=0}^{\infty}\sum_{m=1}^{\infty}x^{a}y^m\frac{1}{m} \left(  \begin{matrix} m+a-1\\ a\\
\end{matrix}\right)t^{m-1}(1+t)^{-(a+m)}\mu^{*m}(m+a-1)\,.
\end{eqnarray*}
Hence  $g_t$ coincides with the generating function of the concentrations that appear in the statement, and the proof can be completed by an appeal to Lemma \ref{L2}(i). 
\QED

We now conclude this section describing the limiting behavior of concentrations as time tends to infinity, starting with the case
$D>0$ (i.e. at the initial time, the mean number of arms is less than the total concentration). 

\begin{corollary}\label{C1} Let $\mu=(\mu(a))_{a\in\N}$ be a probability measure on $\N$
 with mean 
$$\sum_{a\in\N}a\mu(a)=1-D<1\,.$$
Then the solution $(c_t(a,m): a\in\N, m\in\N^*
\hbox{ and } t\geq0)$ to the system \eqref{EQ5}  started from
$$c_0(a,m)={\bf 1}_{\{m=1\}}\mu(a)\,,\qquad a\in\N \hbox{ and } m\in\N^*$$
has a limit as $t\to\infty$ in $\ell^1(\N\times \N^*)$ given by
$$c_{\infty}(a,m)= {\bf 1}_{\{a=0\}}\frac{D}{m}\mu^{*m}(m-1)\,.$$
 \end{corollary}
 
\proof The pointwise convergence (i.e. with $a$ and $m$ fixed) should be plain from the expression given in Theorem \ref{T1}(ii). Recall from Lemma \ref{L1}
that the total concentration $C_t$ tends to $D$ as $t\to \infty$.
On the other hand, as $\mu$ is a probability measure on $\N$ with mean less than $1$, it is easily seen (see e.g. the remark after this proof) that $m^{-1}\mu^{*m}(m-1)$ also defines a probability measure on $\N^*$.
So
$$ \sum_{a=0}^{\infty}\sum_{m=1}^{\infty}c_{\infty}(a,m)=D\,,$$
and we can complete the proof invoking Scheff\'e's lemma (see for example \cite{Billing}). \QED

 It is well-known and easy to check  that if $\mu$ is a probability measure on $\N$ with mean $1-D<1$, then  $\breve{\mu}:=(m^{-1}\mu^{*m}(m-1))_{m\in\N^*}$ is a probability measure on $\N^*$ with mean
$$\sum_{m=1}^{\infty}\mu^{*m}(m-1)=D^{-1}\,.$$
Indeed, we know from the Lagrange inversion formula (see for instance Theorem 5.1.1 in \cite{Wilf}) that the generating functions $g$ of $\mu$ and $h$ of $\breve{\mu}$ are related by the equation $h(x)=xg(h(x))$. The condition on
the mean of $\mu$ reads $g'(1)=1-D<1$; this readily entails that $h(1)=1$, so that $\breve{\mu}$ has total mass $1$. Then taking the (left) derivative at $x=1$, we get
$h'(1)=g(h(1))+ h'(1)g'(h(1))$, which yields $h'(1)=1/D$.
Thus Corollary \ref{C1} implies that
$$\sum_{m=1}^{\infty}mc_{\infty}(0,m)=1\,.$$
This property can be interpreted as follows. Suppose that at the initial time, we tag 
a monomer uniformly at random. Then the distribution of the polymer
at time $t$ that contains this tagged particle
has the distribution
$$\sum_{a=0}^{\infty}\sum_{m=1}^{\infty}m c_t(a,m)\delta_{(a,m)}\,,$$
and as time passes, this family of measures remains tight. Physically, this means that 
{\it if at the initial time  the mean number of arms is less than the total concentration of monomers, then the oriented model does not produce particles with infinite size as time tends to infinity.}

\noindent {\bf Remark. } The recent paper \cite{BSV} deals with a system of randomly interacting particles which is closely related to the present deterministic model, and sheds a probabilistic light on Corollary \ref{C1}. More precisely, in \cite{BSV}, time is discrete and at the initial time 
there are $n$ particles with arms such that the sequence $\xi_1, \ldots, \xi_n$ of the number of arms of particles is i.i.d. with a fixed distribution $\mu$ on $\N$. 
Arms are enumerated uniformly at random, which specifies the order of activation. When an arm is activated, it grabs uniformly at random one of the particles in the system which had not been grabbed previously and which does not belong to its own cluster either. The polymerization procedure terminates when all arms have been activated, and the terminal configuration is given by a forest of  trees.
Roughly speaking, the main result in \cite{BSV} is that if $\mu$ is subcritical (i.e. its first moment is $1-D<1$), then as $n\to\infty$, the distribution
of a tree picked uniformly at random in the terminal configuration converges to that of a Galton-Watson tree with reproduction law $\mu$. According to Dwass \cite{Dwass}, the probability that the size of a Galton-Watson tree with reproduction law $\mu$ is $m$
equals $m^{-1}\mu^{*m}(m-1)$. Further, by the law of large numbers, the number of trees in the terminal forest is approximately $Dn$, so that the density of trees with size $m$ at the terminal time is $Dm^{-1}\mu^{*m}(m-1)$. This corroborates  Corollary \ref{C1}, proving another example of the deep connexions between coagulation models and branching processes (see, e.g. \cite{DT, Spouge'}).

\subsection{Examples} 

Let us now discuss some explicit examples. 

Consider first the degenerate case when $\mu = \delta_1$ is the Dirac mass at $1$, so at the initial time, there is a unit concentration of  particles with unit size having exactly one arm, and all the other concentrations are $0$. It is clear that coagulations then always produce polymers with exactly one arm, and therefore in this specific situation,  the notion of arms plays no role in the evolution of the system. In particular it is not surprising that a known solution should emerge, however it may be interesting to discuss this case as a verification of our general formulas. Specifically we have $ \mu^{*m}=\delta_m$
so we find 
$$c_t(1,m)= t^{m-1}(1+t)^{-(1+m)}\,,$$
and $c_t(a,m)=0$ for $a\neq 1$.
As a check, observe that summing these quantities for $m\in\N^*$ is in
agreement with Lemma \ref{L2}. 
It is also interesting to compare with Smoluchowski's solution \eqref{EQ2}. Specifically
we see that  Smoluchowski's coagulation equation for the kernel $\kappa(m,m')=2$
and monodisperse  initial condition can be viewed as the present oriented coagulation with arms when at the initial time all particles have unit size and exactly one arm.
This observation can be established directly by an elementary analysis of the transition rates in both models.

 More generally, consider the case when $\mu$ is the binomial law with
 parameter $(n,1/n)$, where $n\geq 2$ is some integer, i.e.
 $$\mu(a)= \left(  \begin{matrix} n\\ a\\
\end{matrix}\right)n^{-n}(n-1)^{n-a}\qquad \hbox{for }a=0, \ldots, n\,.$$
 Then $\mu^{*m}$ is the binomial distribution with parameter $(mn,1/n)$, and we get
$$c_t(a,m)=
\frac{(mn)!}{(mn+1-a-m)!m! a!} t^{m-1}(1+t)^{-(a+m)} 
n^{-mn}(n-1)^{mn-a-m+1}
$$
when $a+m-1\leq mn$, and $c_t(a,m)=0$ otherwise.

Then we let $n$ tend to $\infty$ and thus consider the case when $\mu$ is the standard Poisson law, i.e. $\mu(a)=1/( a! \e)$ for $a\in\N$. Then
$\mu^{*m}$ is the Poisson distribution with parameter $m$ and we get
$$c_t(a,m)= \e^{-m}
m^{a+m-1}t^{m-1}(1+t)^{-(a+m)}  \frac{1}{a! m!}\,,\qquad m\in \N\,.$$
It is interesting to point out that summing this quantity over $a\in\N$ yields the total concentration of particles with size $m$,
$$C_t(m):=\sum_{a\in\N}c_t(a,m)=\frac{1}{1+t}\, \frac{(tm/(1+t))^{m-1}}{m!}\e^{-mt/(1+t)}
= \frac{1}{1+t} B(t/(1+t),m)\,. $$
The comparison with Golovin's solution \eqref{EQ3} suggests that, loosely speaking,
 Smoluchowski's coagulation equation for the additive kernel and monodisperse  initial condition coincides after the logarithmic time change $t\to \log(1+t)$, with the present oriented coagulation with arms when at the initial time, particles have unit size and the number of arms is distributed according to the Poisson law.
 
  We leave to the interested reader the task of developing similar calculations for binomial or Poisson laws with mean $\neq 1$ to illustrate Theorem \ref{T1}(ii). 
Here is a final example in this vein. We assume that $\mu$ is the Negative Binomial distribution with parameters $r>0$ and $p\in(0,1)$, viz.
$$\mu(a)=\frac{\Gamma(r+a)}{a!\Gamma(r)} p^r (1-p)^a\,,\qquad a\in\N\,.$$
Recall that 
$$\sum_{a=0}^{\infty}a\mu(a)=r(1-p)/p:=1-D\,,$$
and that $\mu^{*m}$ is Negative Binomial distribution with parameters $mr$ and $p$.
Assuming that $D\neq 0$, we then find that the concentration $c_t(a,m)$ is given by
 $$\e^{Dt}m^{-1} D^{a+1}\left(\e^{Dt}-1\right)^{m-1}
\left(\e^{Dt}-1+D\right)^{-(a+m)}\frac{\Gamma(mr+a+m-1)}{a!(m-1)!\Gamma(rm)} p^{mr} (1-p)^{a+m-1}
\,.$$
\end{section}

\begin{section}{The symmetric model}

\subsection{Setting and relation to a quasi-linear PDE}
We next turn our attention to the symmetric model of coagulation with arms, keeping the notation for the oriented one.
This means that now each aggregation event consumes two arms, one for each particle involved, and that any pair of arms is activated at the same unit rate.
So the transition
\begin{equation}\label{EQ8'}
\{(a,m),(a',m')\}\longrightarrow (a+a'-2,m+m')
\end{equation}
occurs at time $t$ with intensity
$$aa'c_t(a,m)c_t(a',m')\,,$$
and the evolution of the concentration functions is specified by the infinite non-linear system of ODE's
\begin{eqnarray}\label{EQ8}
\d c_t(a,m)  =& &\frac{1}{2}\sum_{a'=1}^{a+1}\sum_{m'=1}^{m-1} a'(a-a'+2)
c_t(a',m')c_t(a-a'+2,m-m')\nonumber \\
&-&c_t(a,m)\sum_{a'=1}^{\infty}\sum_{m'=1}^{\infty}aa'c_t(a',m').
\end{eqnarray}
Again \eqref{EQ8} only makes sense as long $A_t<\infty$, but it is readily seen that $t\to A_t$ decreases as $t$ grows, so it suffices to require that $A_0<\infty$.

Before starting the analysis, let us point at the special role of particles with no arms. Indeed, particles with no arms are inactive in the symmetric model, in the sense that they cannot coagulate with other particles (this was not the case in the oriented model as a particle with no arm could  still be grabbed by some other particle). 
Analytically, this is seen from the fact that the sub-system \eqref{EQ8} for $(a,m)\in\N^*\times \N^*$ is autonomous. 
We also stress that  particles with no arms are produced by the coagulation of two particles both with a single arm, and more precisely,  specializing  \eqref{EQ8}
yields the simple identity
\begin{equation}\label{EQ9}
\d c_t(0,m)  =\frac{1}{2}\sum_{m'=1}^{m-1} 
c_t(1,m')c_t(1,m-m')\,.
\end{equation}

Just as for the oriented model, it is convenient to re-express the system \eqref{EQ8} 
 as
\begin{equation}\label{EQ10}
\d \langle c_t,f\rangle=\frac{1}{2}
\sum_{a,a'=1}^{\infty}\sum_{m,m'=1}^{\infty} (f(a+a'-2,m+m')- f(a,m)-f(a',m')) aa'
c_t(a,m)c_t(a',m')\,,
\end{equation}
where $f:\N\times \N^*\to \R_+$ is a generic  nonnegative and bounded function.

The equations \eqref{EQ8} resembles Smoluchowski's coagulation equation \eqref{EQ1} for the multiplicative kernel. In the latter case, it is well-known that a phenomenon of gelation occurs, in the sense that the total mass is not a preserved quantity for all times
as one might expect naively. Informally, this is due to the formation of particles of infinite size  in finite time; see e.g. \cite{Aldous, Jeon, vDE2}.
A similiar phenomenon may (or may not) happen in the present case, and we shall study \eqref{EQ8} and \eqref{EQ10} before that critical time. Specifically, we
introduce for any $r>0$
$$\Gamma_r:=\inf\{t\geq 0 : \langle c_t, a^2\rangle \geq r\}\,,$$
where by a slight abuse in notation, we write $a^2$ for the function $(a,m)\to a^2$,
and then 
$$\Gamma_{\infty}:=\sup\{\Gamma_r: r\geq 0\}\,.$$ 
Recall also that $A_t:=\langle c_t,a\rangle$ is the mean number of arms at time $t$.

\begin{lemma}\label{L4} For the symmetric model, we have
$$A_t=\frac{A_0}{1+tA_0}\qquad \hbox{for all }t<\Gamma_{\infty}\,.$$
\end{lemma} 

\proof The argument is similar to that in Lemma \ref{L1}. Using the same notation as there, we first specify \eqref{EQ10}  to the function
$f^{(\alpha)}$ and let $\alpha\to \infty$. The bound 
$$|f^{(\alpha)}(a+a'-2,m+m')-f^{(\alpha)}(a,m)-f^{(\alpha)}(a',m')|\leq (a\wedge a')+2\,,$$
enables us to apply the theorem of dominated convergence provided that $\sup_{0\leq s\leq t}\langle c_s, a^2 \rangle <\infty$ and we get the equation
$ \d A_t=-A_t^2$. \QED

We stress that the formula in Lemma \ref{L4} may fail when $t$ is too large. Indeed, 
applying \eqref{EQ10} to $f(a,m)\equiv 2$, we also get $ \d (2C_t)=-A_t^2$
without  requiring that $t<\Gamma_{\infty}$. Thus if Lemma \ref{L4} was always valid,
then the difference $2C_t-A_t$ would remain constant. But this is absurd when
$2C_0<A_0$ since then one would have $2C_t=2C_0+A_t-A_0\to 2C_0-A_0<0$ as $t\to \infty$. 

For the sake of simplicity, we shall focus in the rest of this section on the case when
$A_0:=\langle c_0,a\rangle=1$, which induces no significant loss of generality as the general case
can be reduced to that one by a linear time-change (provided that of course $A_0<\infty$). Recall the notation $g_t(x,y)$ for the generating function of $c_t$ and introduce 
$$k_t(x,y):=\p g_t(x,y)\,,\qquad x,y\in[0,1]$$
which should be viewed as the generating function :
$$k_t(x,y)=\sum_{a=0}^{\infty}\sum_{m=1}^{\infty} x^ay^m (a+1) c_t(a+1,m)\,. $$
The following statement is a partial counter-part of Lemma \ref{L2} for the symmetric model.

\begin{lemma}\label{L5} Assume that $A_0=1$. Then for any solution to the system \eqref{EQ8} we have the equation  
$$\d k_t(x,y)=\left(k_t(x,y)-x A_t\right)\p k_t(x,y)-A_t k_t(x,y)\,,$$
and for  $t< \Gamma_{\infty}$, the latter can be rewritten as the quasi-linear PDE
$$\d k_t(x,y)=\left(k_t(x,y)-\frac{x}{1+t}\right)\p k_t(x,y)-\frac{1}{1+t}k_t(x,y)\,.$$
\end{lemma}

\proof  
Just as in the proof of Lemma \ref{L2}, the equation \eqref{EQ10} specialized to 
$f(a,m)=f_{x,y}(a,m)=x^a y^m$ yields
$$\d g_t(x,y)=\frac{1}{2}
\left(\p g_t(x,y)\right)^2 -x A_t \p g_t(x,y)\,. $$
Then taking the partial derivative with respect to $x$, we obtain the first equation in the statement. 
The second follows from an application of Lemma \ref{L4} and the assumption $A_0=1$.
\QED

We observe that the second PDE in Lemma \ref{L5}  is the 
same as in Lemma \ref{L2}(i) for the oriented model
with $k_t(x,y)$ replacing $g_t(x,y)$. In this direction, we recall that a similar relation between solutions of Smoluchowski's coagulation equations for the additive and the multiplicative kernels holds, see e.g. Theorem 3.9 in  \cite{DT}. 

\subsection{Explicit solutions and examples}
We are now able to solve \eqref{EQ8} up-to the critical time $\Gamma_{\infty}$ when at the initial time, all particles are monomers, i.e. each particle has unit size and its number of arms is arbitrary. Just as for the oriented model, it may be interesting to observe that  the re-parameterization $\alpha=a-2$ is additive on coagulation, in the sense that
\eqref{EQ8'} then reads
\begin{equation}\label{EQ8"}
\{(\alpha,m),(\alpha',m')\}\longrightarrow (\alpha+\alpha',m+m')
\end{equation}
This also makes the solvability of \eqref{EQ5} easier to understand; note however that the parameter $\alpha$ may take negative values and that the rate of the coagulation \eqref{EQ8"} is now
$aa'=(\alpha+2)(\alpha'+2)$ and hence not multiplicative in the alternative parameter $\alpha$.

We consider a measure 
$\mu=(\mu(a))_{a\in\N}$ on $\N$ with unit mean
and finite second moment (we stress that we do not require $\mu$ to be a probability measure), and introduce the probability measure $\nu=(\nu(a))_{a\in\N}$ given by
$$\nu(a)=(a+1)\mu(a+1)\,,\qquad a\in\N\,.$$
We denote  the first moment of $\nu$ by 
$$M:=\sum_{a=0}^{\infty} a \nu(a)=\sum_{a=1}^{\infty}a(a-1)\mu(a)\,,$$
and then define
$$T =\left\{  \begin{matrix} \infty &\hbox{ if }M\leq 1\\ 
1/(M-1) &\hbox{ otherwise.} \\
\end{matrix}\right. $$
Recall also that we write $\nu^{*m}$ for the $m$-th convolution product of $\nu$.

\begin{theorem}\label{T2} 
The system \eqref{EQ8} has a unique solution 
$(c_t(a,m): a, m\in\N^* \hbox{ and } t<T)$  started from
$$c_0(a,m)={\bf 1}_{\{m=1\}}\mu(a)\,,\qquad a,m\in\N^*$$
which is given for $a,m\geq 1$ by
$$c_t(a,m)=\frac{(a+m-2)!}{a! m!}  t^{m-1}(1+t)^{-(a+m-1)}\nu^{*m}(a+m-2)\,.$$
Further, $T$ coincides with the critical time $\Gamma_{\infty}$.
\end{theorem}

\proof 
The initial conditions invite us to introduce the generating function of $\mu$, 
$$g(x):=\sum_{a=0}^{\infty} x^a\mu(a)\,,\qquad x\in[0,1]\,,$$
and to set for $y\in[0,1]$
$$g_0(x,y)=yg(x)\quad \hbox{and} \quad
k_0(x,y):=\p g_0(x,y)=y\sum_{a=0}^{\infty} x^a\nu(a).$$

Consider the second PDE of Lemma \ref{L5}, viz. 
\begin{equation}\label{EQ11}
\d k_t(x,y)=\left(k_t(x,y)-\frac{x}{1+t}\right)\p k_t(x,y)-\frac{1}{1+t}k_t(x,y)\,,
\end{equation}
with boundary value $k_0(x,y)$ defined above. If we replace $k_t$ by $g_t$, this
is precisely  the PDE that has been solved by the method of characteristics  in Lemma \ref{L2}(i). There is however a difference that requires some attention : here we assume that
$\p k_0(1,1)=M<\infty$ whereas we had the stronger hypothesis $\p g_0(1,1)=1$ in Lemma \ref{L2}(i). Nonetheless, the condition  $t<T$ implies $1+t-t M>0$, and thus
the derivative of the function $x\to (1+t)x-tk_0(x,y)$ is strictly positive for $x\in[0,1]$. As
$-tk_0(0,y)<0<1\leq 1+t-tk_0(1,y)$, this ensures the existence of a unique inverse function
$\ell_t(\cdot,y):[0,1]\to[0,1]$ to $x\to (1+t)x-tk_0(x,y)$. 
The argument in the proof of Lemma \ref{L2}(i) is thus still valid, and we conclude that
for every $t<T$  and $x,y\in[0,1]$
$$k_t(x,y) = (1+t)^{-1}k_0(\ell_t(x,y),y)= t^{-1} \ell_t(x,y)-\frac{x}{t^2+t}\,.$$

The comparison with Theorem \ref{T1}(i) now yields that for $t<T$ and $x,y\in[0,1]$
$$k_t(x,y)=\sum_{a=0}^{\infty}\sum_{m=1}^{\infty}x^ay^m
 m^{-1}t^{m-1}(1+t)^{-(a+m)} \left(  \begin{matrix} a+m-1\\ a\\
\end{matrix}\right) \nu^{*m}(a+m-1)\,.$$
We thus see that if we define $c_t(a,m)$ for $a,m\in\N^*$ and $0\leq t< T$ as in the statement, and let $k_t$ be the generating function of $((a+1)c_t(a+1,m): a\in\N \hbox{ and }m\in\N^*)$, then
$k_t$ solves \eqref{EQ11} for $0\leq t<T$. 

We next observe that $\ell_t(1,1)=1$ since $k_0(1,1)=1$, and thus
$$k_t(1,1)=\langle c_t, a\rangle =1/(1+t)\,.$$
Hence $k_t$ also solves
$$\d k_t(x,y)=\left(k_t(x,y)-x\langle c_t, a\rangle\right)\p k_t(x,y)-\langle c_t, a\rangle k_t(x,y)\,,$$
which is the first PDE in Lemma \ref{L5}. 
Inverting the generating functions, we conclude that $(c_t(a,m): a,m\in\N^*\hbox{ and }
0\leq t < T)$ is a solution to \eqref{EQ8}.  

We then check that the critical time
$\Gamma_{\infty}$ of this solution coincides with $T$. 
In this direction, we first recall that $\p k_0(1,1)=M$, $\ell_t(1,1)=1$ and, by definition, that
$$(1+t)\ell_t(x,1)=tk_0(\ell_t(x,1),1)+x\,.$$
We take the (left) derivative with respect to the variable $x$ at $x=1$
and obtain 
$$(1+t)\p \ell_t(1,1)=tM\p \ell_t(1,1) + 1\,,$$
so 
$$\p \ell_t(1,1)=\frac{1}{1+t(1-M)}\,,\qquad t<T\,.$$
It follows that
$$\p k_t(1,1) = t^{-1}\p \ell_t(1,1)-(t+t^2)^{-1}=\frac{M}{(1+t)(1+t(1-M))}$$
remains bounded on compact intervals in $[0,T[$, and further explodes as $t \uparrow T$ when $M>1$ (i.e. $T<\infty$). As $\p k_t(1,1)=\langle c_t, a^2-a\rangle$, 
we conclude that $T=\Gamma_{\infty}$. Finally, the uniqueness of the solution to \eqref{EQ8} up to the critical time should be plain from a perusal of the preceding arguments.
 \QED

Theorem \ref{T2} does not provide an expression for concentrations of particles with no arms. However, these can be recovered from the concentrations of particles with exactly one arm via the equation \eqref{EQ9} (note that the concentration of monomers with no arm does not evolve, so we may focus on particles with size $m\geq 2$ in the following statement). 

\begin{corollary}\label{C2} Under the same assumptions as in Theorem \ref{T2}, and assuming also that $\nu(0)>0$,
we have for every $m\geq 2$ and $t<T$
$$c_t(0,m)=\frac{1}{m(m-1)}(1+1/t)^{1-m}\nu^{*m}(m-2)\qquad \hbox{ for }m\geq 2\,.$$

As a consequence, in the case when $M\leq 1$ (that is when $T=\infty$), the concentration of particles $(a,m)$ at time $t$ has a limit $c_{\infty}(a,m)$
as $t\to \infty$ which is given by
$$c_{\infty}(a,m)= {\bf 1}_{\{a=0\}}\frac{1}{m(m-1)}\nu^{*m}(m-2)\qquad \hbox{ for }
a\in\N \hbox{ and }m\geq 2\,.$$
\end{corollary}

\proof We know from Theorem \ref{T2} that for every $t<T$ and $m\geq 1$
$$c_t(1,m)=t^{m-1}(1+t)^{-m}m^{-1}\nu^{*m}(m-1)\,.$$
According to the classical Lagrange inversion formula (cf. Theorem 5.1.1 in \cite{Wilf}),
$m^{-1}\nu^{*m}(m-1)$ appears as the $m$-th coefficient in the analytic expansion 
of the entire function $u$ which solves $u(x)=x\phi(u(x))$, where $\phi$ is the generating function of the probability measure $\nu$ (the assumption $\nu(0)>0$ ensures that
$\phi(0)>0$).
Another application of the Lagrange inversion formula now shows that
 the $m$-th coefficient in the analytic expansion of the entire function $u^2(x)$
 is $2m^{-1}\nu^{*m}(m-2)$. In terms of convolution product, this reads
$$c_t(1,\cdot)^{*2}(m): =\sum_{m'=1}^{m-1} 
c_t(1,m')c_t(1,m-m')=
 t^{m-2}(1+t)^{-m} \frac{2}{m}\nu^{*m}(m-2)\,.$$
Since $c_0(1,m)=0$ for $m\geq 2$ by assumption and
$$\int_0^t  s^{m-2}(1+s)^{-m}{\rm d} s
=(m-1)^{-1}(1+1/t)^{1-m}\,,$$
our first claim follows from \eqref{EQ9}. The second follows immediately from the first for $a=0$, and from Theorem \ref{T2} for $a\geq 1$. \QED

 We stress that when $M\leq 1$ and $\nu\neq \delta_1$, then
\begin{equation}\label{EQ12}
\sum_{a=0}^{\infty}\sum_{m=2}^{\infty}m c_{\infty}(a,m)=\sum_{m=2}^{\infty}\frac{1}{(m-1)}\nu^{*m}(m-2)
=\sum_{a=0}^{\infty}\frac{1}{a+1}\nu(a)=\mu(\N^*)\,.
\end{equation}
Just as for the oriented model, this identity can be interpreted as a tightness property 
  as time passes 
for the distribution of the polymer
at time $t$ that contains a randomly tagged monomer. Physically, this means that the symmetric coagulation model does not produce particles with infinite size as time tends to $\infty$.

 Indeed, the identity \eqref{EQ12} is easily checked as follows :
\begin{eqnarray*}
\sum_{m=2}^{\infty}\frac{1}{(m-1)}\nu^{*m}(m-2)&=&\sum_{a=0}^{\infty}\nu(a)\sum_{m=2}^{\infty}\frac{1}{(m-1)}\nu^{*(m-1)}(m-2-a)\\
&=&\sum_{a=0}^{\infty}\nu(a)\sum_{n=1}^{\infty}n^{-1}\nu^{*n}(n-a-1)\,.\\
\end{eqnarray*}
As $\nu\neq \delta_{1}$ is a probability measure with mean $M\leq 1$, it follows easily from the Lagrange inversion formula (see the argument in the remark after the proof of Corollary \ref{C1}) that for any $a\geq 0$, 
$$\sum_{n=1}^{\infty}\frac{a+1}{n}\nu^{*n}(n-a-1)=1\,.$$
We stress that this is where the assumption  $\nu\neq \delta_{1}$ is needed as otherwise the generating function of $\nu$ would vanish at $0$, impeding the application of Lagrange inversion formula. We thus have checked that
$$\sum_{m=2}^{\infty}\frac{1}{(m-1)}\nu^{*m}(m-2)=\sum_{a=0}^{\infty}(a+1)^{-1}\nu(a)\,,$$
the other equalities in  \eqref{EQ12} are obvious.

To summarize this observation, if we exclude the case when $\nu=\delta_1$
(or equivalently, $\mu=\frac{1}{2}\delta_2$) that will discussed further below, we have shown that there either gelation occurs at a finite time (by  Theorem \ref{T2}, this happens if and only if $M> 1$), or the mass distribution of particles remains tight as time tends to infinity. 

\noindent{\bf Remark. } The limiting concentrations in Corollary \ref{C2}  bear a striking
resemblance with the distribution of the total population generated by a (sub)-critical Galton-Watson branching process with reproduction law $\nu$  and started from two ancestors. We refer the interested reader to
the recent work \cite{BS} for a probabilistic interpretation that relies on the study of the random configuration model. The latter can be used to create typical graphs with specified degree sequence, and is constructed via a stochastic algorithm which
can be thought of as the probabilistic counterpart of the present symmetric model of coagulations with limited aggregations.

We now conclude this work by illustrating Theorem \ref{T2} with some examples.
The simplest is when $\mu=\delta_1$, that is, initially, there is a unit concentration of monomers with a single arms. So $\nu=\delta_0$ and one finds
$$c_t(1,1)=\frac{1}{1+t}\quad,\quad c_t(0,2)=\frac{t}{2t+2}\,,\qquad 0\leq t < T=\infty\,,$$
and all the other concentrations are zero.

Next, suppose that at the initial time, there is a concentration $1/2$ of monomers with two arms, i.e. $\mu=\frac{1}{2}\delta_2$. Then $\nu=\delta_1$ and we get
$$c_t(2,m)=\frac{1}{2} t^{m-1}(1+t)^{-(m+1)}\,,\qquad m\in\N^*,0\leq t < T=\infty\,,$$
and all the other concentrations are zero. Note the similarity with the oriented model started from a unit density of monomers with a single arm. Of course, this property could also be observed by a direct argument.

We then turn our attention to the simplest example with a finite critical time, namely the situation where at the initial time, there is a concentration $1/3$ of monomers with three arms, i.e. $\mu=\frac{1}{3}\delta_3$. Then $\nu=\delta_2$ and $M=2$, so the critical time is $T=1$. 
One gets
$$c_t(m+2,m)=\frac{(2m)!}{(m+2)! m!}  t^{m-1}(1+t)^{-(2m+1)}\,,\qquad m\in\N^*\hbox{ and }t<1\,,$$
and all the other concentrations are $0$. It is easily checked  that
$$\lim_{t\to 1-}\sum_{m=1}^{\infty}(m+2)^2c_t(m+2,m)=\infty\,,$$
in agreement with Lemma \ref{L4}. 

Finally, consider the case when $\mu$ is the standard Poisson law, then $\nu$ is also the standard Poisson law and in particular $T=\infty$. As $\nu^{*m}$ is the Poisson law with parameter $m$, we get from Theorem \ref{T2} that
$$c_t(a,m)=\frac{m^{a+m-2}}{a! m!}  t^{m-1}(1+t)^{-(a+m-1)}\e^{-m}\,,\quad a,m\in\N^*, t\geq 0\,,$$
and from Corollary \ref{C2} that 
$$
c_t(0,1)=\e^{-1} \quad \hbox{and}\quad
c_t(0,m) = \e^{-m}\frac{m^{m-2}}{m!}(1+1/t)^{1-m} \hbox{ for }m\geq 2\,.
$$
It is interesting to point out that 
$$\lim_{t\to\infty}c_t(0,m)=\e^{-m}\frac{m^{m-2}}{m!}=m^{-1}B(1,m)\,,\qquad m\in\N^*$$
and to compare with McLeod's solution \eqref{EQ4} to Smoluchowski's coagulation equation for the multiplicative kernel and monodisperse initial condition. 
We see that the terminal state for the symmetric model of coagulation started from monomers with arms distributed according to the standard Poisson law is the same as 
the state at gelation time in Smoluchowski's model for the multiplicative kernel and monodisperse initial condition. 

\end{section}

\end{document}